\documentclass[12pt,a4paper]{article}\usepackage{a4wide}

\usepackage{epsfig}
\usepackage{amssymb,latexsym,amsmath}
\usepackage{amsfonts}
\usepackage{amsbsy}

\numberwithin{equation}{section}

\DeclareMathOperator{\tr}{tr}

\newcommand{\ui}{\textrm{i}}
\newcommand{\ue}{\textrm{e}}
\newcommand{\UI}{\textrm{I}}

\newcommand{\ud}{\mathrm{d}}

\newcommand{\SM}{{\mathbb S}}

\newcommand{\GSE}{\textrm{GSE}}

\newcommand{\CSE}{\textrm{CSE}}
\newcommand{\COE}{\textrm{COE}}
\newcommand{\SU}{\textrm{SU}}

\begin{document}

\thispagestyle{empty}

\noindent ULM-TP/03-3\\April 2003
\vspace{2cm}

\begin{center}

{\LARGE\bf   The spin contribution to the form factor of
\vspace*{2mm} \\ quantum graphs} \\ 
\vspace*{3cm}
{\large Jens Bolte}%
\footnote{E-mail address: {\tt jens.bolte@physik.uni-ulm.de}}
{\large and Jonathan Harrison}%
\footnote{E-mail address: {\tt jon.harrison@physik.uni-ulm.de}}

\vspace*{1cm}

Abteilung Theoretische Physik\\
Universit\"at Ulm, Albert-Einstein-Allee 11\\
D-89069 Ulm, Germany 
\end{center}

\vfill

\begin{abstract} 
Following the 
quantisation of a graph with the Dirac operator (spin-$1/2$)
we explain how additional weights in the spectral form factor $K(\tau)$
due to spin propagation around orbits 
produce higher order terms in the small-$\tau$ asymptotics 
in agreement with symplectic random matrix ensembles.   
We determine conditions on the
group of spin rotations sufficient to generate 
CSE statistics.
\end{abstract}


\newpage

\section{Introduction}\label{s:introduction}
Overwhelming evidence shows that correlations in discrete energy spectra
of classically chaotic quantum systems generally follow the conjecture
of Bohigas, Giannoni, and Schmit \cite{paper:bohigasgiannonischmit}.
According to this the spectral statistics can be described by 
random matrix theory (RMT), the universality classes being completely
determined by symmetry. 
Berry \cite{paper:berry} analysed spectral 
two-point correlations measured with the form factor $K(\tau)$ obtaining the 
leading order in its small-$\tau$ asymptotics in agreement with RMT.
Recent developments  suggest further asymptotic terms can be attributed to 
correlations between classical periodic orbits with (almost)
self-intersections \cite{paper:sieberrichter, paper:sieber}. Implicit
to this method is the assumption that the $\tau^{m}$-term in the form factor 
expansion is determined only by correlations between pairs of orbits where 
the order of sections of the orbit has been changed at $m-1$ (almost)
self-intersections. The original results of Sieber and Richter, concerning 
systems on surfaces of constant negative curvature, were extended to a class 
of graphs quantised with the Schr\"odinger operator by Berkolaiko, Schanz,
and Whitney 
\cite{paper:berkolaikoschanzwhitney, paper:berkolaikoschanzwhitney2}.
They find agreement with the form factor $K_{\COE}(\tau)$ of
the circular orthogonal RMT-ensemble (COE) to third order in $\tau$ by 
considering correlations between orbits with up to two self-intersections. 

In \cite{paper:bolteharrison} we showed  that a graph quantised with 
the Dirac operator and possessing time-reversal symmetry produces level 
statistics which agree numerically with those of the Gaussian symplectic 
ensemble (GSE), in accordance with \cite{paper:bohigasgiannonischmit}.  
The form factor differs from that of the usual Schr\"odinger quantisation 
via weights determined by the spin transformations around classical periodic 
orbits.  In the diagonal approximation we found an agreement of the form
factor with $K_{\GSE}(\tau)$ in first order.  Here we show that the
spin weights generate the additional pre-factors relating the expansions of 
$K_{\CSE}(\tau)$ and $K_{\COE}(\tau)$ in all orders. 
Our approach avoids the various ad hoc assumptions introduced in the related 
investigation by Heusler \cite{paper:heusler} and allows us to state precise 
conditions on the spin dynamics. 
The result also provides strong evidence for the hypothesis that the 
$\tau^{m}$-term derives only from pairs of orbits differing in the order of 
sections at $m-1$ self-intersections.  
\section{The form factor}\label{s:form factor}
In our previous work \cite{paper:bolteharrison} we studied correlations in
spectra of Dirac operators on graphs.  Here we rather adopt the closely
related point of view taken in 
\cite{paper:berkolaikoschanzwhitney, paper:berkolaikoschanzwhitney2}
and consider the form factor derived from the spectrum of the $S$-matrix
that was introduced in \cite{paper:bolteharrison}.  On the side of RMT we
hence have to consider the circular instead of the Gaussian ensembles.  

We recall that a (compact) graph consists of $V$ vertices connected by $B$ 
bonds.  The valency of a vertex $i$ is $v_i$.  Let $(ij)$ label a transition 
from vertex $i$ to vertex $j$ along a bond $\{ ij \}$.  According to 
\cite{paper:bolteharrison} for fixed energy there exist two linearly 
independent eigenspinors of the Dirac operator for each transition.  
The $S$-matrix $\SM$ is the matrix of transition elements connecting the 
eigenspinors.  It is therefore a square matrix of dimension $4B$.  We divide 
the $S$-matrix into $2\times 2$ blocks $\SM^{(ij)(kl)}$ defining transitions 
between the pair of eigenspinors traveling from $k$ to $l$ and a pair 
traveling from $i$ to $j$.  The $S$-matrix is then defined by
\begin{equation}\label{S-matrix}
\SM^{(ij)(kl)}:=\delta_{il}\,  \sigma_{(ij)(ki)} \, u^{(ij)(ki)} 
\, \ue^{\ui \phi_{\{ kl \} }} \ ,
\end{equation}
where $u^{(ij)(ki)}$ is an element of $\SU (2)$ describing the spin 
transformation at the vertex $i$ and the terms $\sigma_{(ij)(ki)}$ define a 
$2v_i \times 2v_i$ unitary matrix  $\Sigma^{(i)}$.  The Kronecker-delta in 
(\ref{S-matrix}) ensures transitions only occur between 
bonds connected at 
a vertex.  The phases $\phi_{\{ kl \} }$ are random variables uniformly 
distributed in $[ 0 , 2 \pi ]$.  They define an ensemble of matrices 
$\SM_{\phi}$ over which we average.  Such an average is equivalent 
to a spectral averaging when considering the level 
statistics of a Dirac operator on the graph.

Time-reversal invariance requires $\Sigma^{(i)}$ to be symmetric and 
\begin{equation}\label{eq:trs}
\SM^{(lk)(ji)}=\bigl| \SM^{(ij)(kl)} \bigr| \left( \SM^{(ij)(kl)} 
\right)^{-1} \ .
\end{equation}  
To satisfy (\ref{eq:trs}) we define spin transformations
\begin{equation}\label{spin transform}
u^{(ij)(ki)}:= u^{(i)}_{j} \big( u^{(i)}_{k} \big)^{-1} \ .
\end{equation}
The $v_i$ elements $u^{(i)}_{j} \in \SU (2)$ define all spin transformations 
at the vertex $i$.  See \cite{paper:bolteharrison} for details.

Having defined the $S$-matrix of a Dirac graph the form factor may be 
introduced as in \cite{book:haake}.  We remove Kramers' degeneracy present
in systems with half-integer spin and time-reversal invariance as in
\cite{paper:boltekeppeler3}. This leaves us with $N=2B$ eigenvalues of $\SM$, 
leading to
\begin{equation}\label{eq:form factor defn}
K(\tau) := \frac{1}{4N} \left\langle | \tr \SM_{\phi}^{n} |^{2} 
\right\rangle_{\phi} \ .
\end{equation}
The trace of $\SM^{n}$ may be expanded as a sum over the set $P_n$ of 
periodic orbits of length $n$, 
\begin{equation}\label{trace expansion}
\tr \SM_{\phi}^n = \sum_{p\in P_n} \frac{n}{r_{p}} \, A_{p} \, 
\ue^{\ui\pi\mu_p} \, \tr (d_{p}) \, \ue^{\ui \phi_{p}} \ ,
\end{equation}
where the periodic orbit $p$ consists of a series of transitions 
$(b_1 , b_2 , \dots , b_n )$ and
\begin{equation}
\begin{split}
A_p \, \ue^{\ui\pi\mu_p} 
       &:=\sigma_{b_n b_{n-1}} \sigma_{b_{n-1} b_{n-2} }
\dots \sigma_{b_2 b_1} \ ,\\
d_p    &:=u^{b_n b_{n-1}} u^{b_{n-1} b_{n-2}} \dots u^{b_2 b_1} \ ,\\
\phi_p & :=\sum_{j=1}^{n} \phi_{\{ b_j \} } \ .
\end{split}
\end{equation}
The phases $\mu_p$ are such that $A_p >0$, and $r_{p}$ is the repetition 
number of $p$ so that $n/r_p$ is the number of possible starting positions 
of an orbit up to cyclic permutations.
Substituting the periodic orbit expansion (\ref{trace expansion})
into (\ref{eq:form factor defn}) and carrying out the average over $\phi$ 
we obtain 
\begin{equation}\label{sym form factor}
K_{\mathrm{sympl}}(\tau) = \frac{n^2}{4(2B)} \sum_{p,q \in P_{n}} 
\frac{A_{p} A_{q}}{r_p r_q} \, \ue^{\ui\pi(\mu_p-\mu_q)} \, 
\tr(d_p) \tr(d_q) \, \delta_{\phi_p ,\phi_q }  
\end{equation}
where $\tau = n/2B$.  We label the form factor $K_{\mathrm{sympl}}$ according 
to the symplectic symmetry introduced by time-reversal invariance in a system 
with spin-$1/2$.  The Kronecker-delta fixes contributing terms in 
$K_{\mathrm{sympl}}$ to pairs of orbits in which each bond is visited the 
same number of times.  On a metric graph with rationally independent bond 
lengths this is equivalent to requiring the lengths of $p$ and $q$ be equal.

For comparison the form factor studied in 
\cite{paper:berkolaikoschanzwhitney, paper:berkolaikoschanzwhitney2}
for a graph quantised with the Schr\"odinger operator (spin-$0$) in a 
time-reversal symmetric fashion is
\begin{equation}\label{orth form factor}
K_{\mathrm{orth}}(\tau) = \frac{n^2}{2B} 
\sum_{p,q \in P_{n}} \frac{A_{p} A_{q}}{r_p r_q} \, \ue^{\ui\pi(\mu_p-\mu_q)}  
\, \delta_{\phi_p ,\phi_q } \ , 
\end{equation}
where the  definition of $A_p$ remains the same. 
This form factor is labeled by the orthogonal 
symmetry of the system.  

It was pointed out in \cite{paper:heusler} that 
comparing the RMT form factors of the CSE and COE makes clear 
the close connection between them,
\begin{equation}
\begin{split}
K_{\CSE}(\tau) & =  \frac{\tau}{2} + \frac{\tau^2}{4} + \frac{\tau^3}{8}
+ \frac{\tau^4}{12} + \dots \ ,\\
\frac{1}{2} K_{\COE}\left( \frac{\tau}{2} \right) 
& =  \frac{\tau}{2} - \frac{\tau^2}{4} + \frac{\tau^3}{8}
- \frac{\tau^4}{12} + \dots \ .
\end{split}
\end{equation}
Calling $K^{m}$ the term containing $\tau^m$ the relationship may be 
written
\begin{equation}
K^{m}_{\CSE}(\tau) = \left( -\frac{1}{2} \right)^{m+1} K^{m}_{\COE}(\tau) \ .
\end{equation}
According to the conjecture of Bohigas, Giannoni, and Schmit 
\cite{paper:bohigasgiannonischmit} in the semiclassical limit we expect the 
form factors of quantum graphs to correspond to those of random matrices.  
In particular,
\begin{equation}\label{form factor relation}
K^{m}_{\mathrm{sympl}}(\tau) = \left( -\frac{1}{2} \right)^{m+1} 
K^{m}_{\mathrm{orth}}(\tau) \ .
\end{equation}
It is this relation we wish to demonstrate in quantum graphs.
\section{Spin contributions to the form factor}\label{s:spin contribution}
The form factor on graphs may be studied analytically in the semiclassical 
limit.  The system is defined by the matrix $\SM$ of dimension $4B$ and so 
the semiclassical limit is $B\rightarrow \infty$.  For small but finite
$\tau=n/2B$ the limit of long orbits, $n\rightarrow \infty$, is also required.
For details see 
\cite{paper:berkolaikoschanzwhitney, paper:berkolaikoschanzwhitney2}.
In this limit the proportion of orbits $p$ with $r_p \ne 1$ 
tends to zero so these orbits can effectively be ignored in equations 
(\ref{sym form factor}) and (\ref{orth form factor}).  Following
\cite{paper:berkolaikoschanzwhitney2} the sum over orbit pairs is organised 
in terms of diagrams.  A diagram consists of all pairs of orbits 
related by the same pattern of 
permutations of arcs between self-intersections
and time-reversal of arcs.   Consequently such pairs of orbits 
have identical phases $\phi_p=\phi_q$.  
Figures \ref{fig:intersection} and 
\ref{fig:pair} provide examples of diagrams.

The contribution to the form factor from a specific diagram $D$ 
with $m-1$ self-intersections is
\begin{equation}\label{eq:sym 1}
K^{m,D}_{\mathrm{sympl}} (\tau) := \frac{n^2}{4(2B)} \sum_{(p,q) \in D_n}
A_{p} A_{q} \, \ue^{\ui\pi(\mu_p-\mu_q)} \, \tr(d_p) \tr(d_q) \ .
\end{equation}
Here $D_{n}$ is the set of pairs of orbits $(p,q)$ of length 
$n$ contained in $D$.

To separate spin contributions from this sum we assume that
the elements $d_p$ are chosen randomly (independent of $p$) from a (sub-) 
group $\Gamma \subseteq \SU(2)$.  This can be achieved by selecting the 
elements $u^{(i)}_j$ randomly from $\Gamma$.  Then
\begin{equation}\label{eq:sym 2}
K^{m,D}_{\mathrm{sympl}} (\tau) = \frac{1}{4} 
\left( \frac{1}{| D_{n}|} \sum_{(p,q) \in D_{n}} \tr(d_p) \tr(d_q) \right)
\times \left( \frac{n^2}{2B}
\sum_{(p,q) \in D_{n}} A_{p} A_{q} \, \ue^{\ui\pi(\mu_p-\mu_q)} \right) \ .
\end{equation}
The second term is the equivalent contribution to the 
orthogonal form factor (\ref{orth form factor}),
\begin{equation}\label{eq:sym 3}
K^{m,D}_{\mathrm{orth}}(\tau)  :=  \frac{n^2}{2B}
\sum_{(p,q) \in D_{n}} A_{p} A_{q} \, \ue^{\ui\pi(\mu_p-\mu_q)} \ .
\end{equation}
We will show that in the semiclassical limit
\begin{equation}\label{eq:result}
\langle \tr(d_p) \tr(d_q) \rangle^{m,D_n} := 
\frac{1}{| D_n|} \sum_{(p,q) \in D_n} \tr(d_p) \tr(d_q) 
\to \left( - \frac{1}{2} \right)^{m-1} \ ,
\end{equation}
independent of the diagram $D$ for fixed $m$.  
Substituting into (\ref{eq:sym 2}) generates 
the relation (\ref{form factor relation}) between the orthogonal and 
symplectic form factors.  
At this point we remark that the spin contribution to the form factor defined 
in (\ref{eq:result}) is different from that in \cite{paper:heusler}.  

To determine the spin contributions we first take the case $\Gamma=\SU(2)$, 
ie random spin rotations are chosen from the whole of $\SU(2)$ with Haar 
measure.  Each arc of $p$ contributes a random element of $\SU(2)$ 
to $d_p$.  If $(p,q) \in D$ in $d_q$ the order of the product is 
changed and some elements are
replaced with their inverse.  In the semiclassical limit where
the number of orbits tends to infinity the sum over pairs of orbits may be 
replaced by integrals over $\SU(2)$ for each arc of the diagram. To evaluate 
the spin contributions we require three identities:
\begin{eqnarray} \label{identity 1}
\int_{\SU(2)} \tr(xuyu) \ \ud u &=& -\frac{1}{2} \tr(xy^{-1}) 
\\ \label{identity 2}
\int_{\SU(2)} \tr(xuyu^{-1}) \ \ud u &=& \frac{1}{2} \tr(x) \tr(y) 
\\ \label{identity 3}
\int_{\SU(2)} \tr(xu)\tr(yu) \ \ud u &=& \frac{1}{2} \tr(xy^{-1}) 
\end{eqnarray} 
where $u,x,y \in\SU(2)$.  Equations (\ref{identity 1})--(\ref{identity 3}) 
may be evaluated directly by parameterising $\SU(2)$.
To establish (\ref{eq:result}) we consider changing the order of arcs at 
one self-intersection or between a pair of intersections.  All diagrams can 
be generated via these operations.  Counting the number of intersections 
at which the order of elements is changed then proves the result. 
\subsection{Reordering at a single intersection}\label{ss:single}
\begin{figure}[htb]
\begin{center}
\includegraphics[width=9cm]{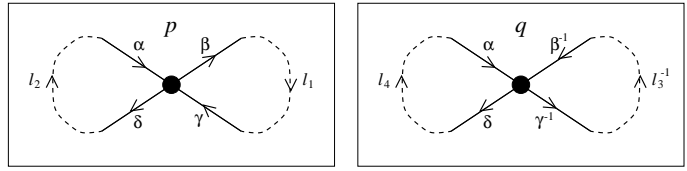}
\end{center}
\caption{A pair of orbits $(p,q)$ with the order of arcs changed at a single 
self-intersection}
\label{fig:intersection}
\end{figure}

Figure \ref{fig:intersection} shows two alternative orders at a single 
intersection.  For a pair of orbits with such a self-intersection
the spin contribution in the semiclassical limit is
\begin{equation}
\langle \tr(d_p) \tr(d_q) \rangle = 
\int_{\SU(2)}\dots\int_{\SU(2)} \tr(\alpha \beta l_1 \gamma \delta l_2)
\, \tr(\alpha \gamma^{-1} l_3^{-1} \beta^{-1} \delta l_4) \ \ud \alpha \,
\ud \beta \, \ud \gamma \, \ud \delta \dots
\end{equation}
Changing variables so $x:=\alpha \beta$ and $y:=\gamma \delta$ and using 
(\ref{identity 1}) we obtain
\begin{equation}
\langle \tr(d_p) \tr(d_q) \rangle = 
-\frac{1}{2} \int_{\SU(2)}\dots\int_{\SU(2)} \tr( x l_1 y l_2)
\, \tr(x l_3 y l_4) \, \ud x \,\ud y \dots
\end{equation}
Reordering terms in $d_q$ at one vertex therefore introduces a factor of 
$-1/2$ to the spin contribution.
  
\subsection{Reordering at a pair of intersections}\label{ss:pair}
\begin{figure}[htb]
\begin{center}
\includegraphics[width=7cm]{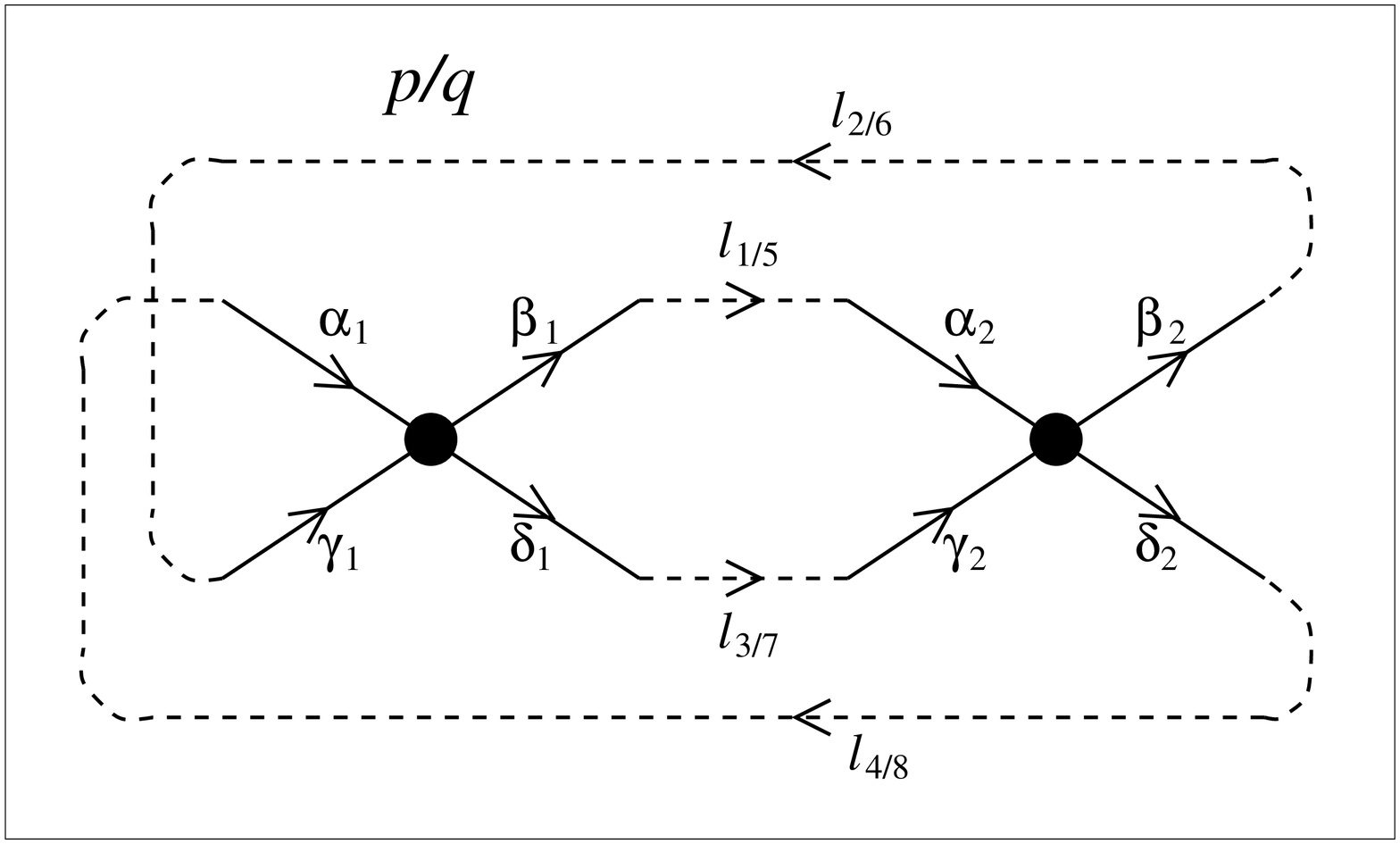}
\end{center}
\caption{Reordering arcs between a pair of self-intersections}
\label{fig:pair}
\end{figure}

The procedure in \ref{ss:single} uses the time-reversal invariance
of the system to reverse the directions of arcs.  There is a second type 
of reordering of arcs independent of this symmetry which is possible when 
multiple arcs run between a pair of self-intersections, figure 
\ref{fig:pair}.  The sections $\beta_1 \dots \alpha_2$ and 
$\delta_1 \dots \gamma_2$ can then be taken in either order.  
The relevant spin 
contribution is
\begin{equation}\label{eq:pair}
\begin{split}
\langle \tr(d_p) \tr(d_q) \rangle  = &
\int_{\SU(2)}\dots\int_{\SU(2)} \tr(\alpha_1 \beta_1 l_1 \alpha_2 \beta_2 
l_2 \gamma_1 \delta_1 l_3 \gamma_2 \delta_2 l_4) \,
\tr(\alpha_1 \delta_1 l_7 \gamma_2 \beta_2 l_6 
\gamma_1 \beta_1 l_5 \alpha_2 \delta_2 l_8) \times \\ &
\hspace*{2.5cm} \times \ud \alpha_1 \, \ud \alpha_2 \, \ud \beta_1 \, 
\ud \beta_2 \, \ud \gamma_1 \, \ud \gamma_2 \, \ud \delta_1 \, \ud \delta_2 
\dots \ .
\end{split}
\end{equation}
We notice that exchanging the order of the central arcs changes the order of
elements of $\SU(2)$ at both self-intersections.  
From (\ref{identity 2}) and (\ref{identity 3}) it can be shown that
\begin{equation}
\int_{\SU(2)}\int_{\SU(2)} \tr(uav^{-1} b u^{-1} c v d) \ \ud u \, \ud v = 
\frac{1}{4} \tr (cbad) \ .
\end{equation}
To apply this to (\ref{eq:pair}) make
substitutions 
$x_j:=\alpha_j \beta_j , \, y_j:=\gamma_j \delta_j$, 
then
\begin{equation}
\begin{split}
\langle \tr(d_p) \tr(d_q) \rangle  = 
\frac{1}{4} \int_{\SU(2)} \dots \int_{\SU(2)}
&\tr(x_1 l_1 x_2 l_2 y_1 l_3 y_2 l_4) \,
 \tr(x_1 l_5 x_2 l_6 y_1 l_7 y_2 l_8) \times \\
&\qquad\times\ud x_1 \, \ud x_2 \, \ud y_1 \, \ud y_2 \dots \ .
\end{split}
\end{equation}
Reordering by exchanging arcs of a diagram introduces a factor of $1/4$ 
to the spin contribution and requires changing the order at two 
self-intersections simultaneously.  

\subsection{Counting self-intersections}\label{ss:counting}
\begin{figure}[htb]
\begin{center}
\includegraphics[width=3cm]{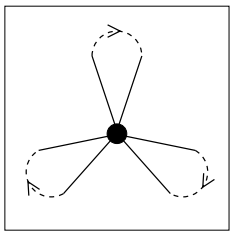}
\end{center}
\caption{Orbit with degenerate self-intersection}
\label{fig:loops}
\end{figure}

All diagrams can be constructed via combinations of the procedures described 
in sections \ref{ss:single} and \ref{ss:pair}.  This is not obvious, for 
example a system of loops in which the same self-intersection is visited 
twice does not appear to fall within this classification, 
figure \ref{fig:loops}.  
In fact degenerate self-intersections allow both types of 
reordering at the 
intersection.  To distinguish the cases it is necessary to follow the orbit 
counting each intersection when it is reached after determining if the order 
of arcs at the intersection has been changed.  
The number of self-intersections for a given diagram is then $m-1$ (note our 
multiple counting of degenerate self-intersections differs from the definition
in \cite{paper:berkolaikoschanzwhitney2}).  As each intersection 
effectively contributes a factor $-1/2$ we obtain 
\begin{equation}\label{eq:result 2}
\langle \tr(d_p) \tr(d_q) \rangle^{m,D_n}  
\rightarrow \left( -\frac{1}{2} \right)^{m-1} \int_{\SU(2)} 
\big( \tr(u)  \big)^2 \ud u \ .
\end{equation}
The final integral over $\SU(2)$ is
\begin{equation}
\int_{\SU(2)} \big( \tr(u) \big)^2 \ud u = 1 
\end{equation}
as the defining representation of $\SU(2)$ is naturally irreducible.

We remark that if diagrams with $m'-1 \ne m -1$ self-intersections contributed 
to $K^{m}_{\textrm{orth}}(\tau)$ in such a way that nevertheless 
$K_{\textrm{orth}}(\tau)=K_{\COE}(\tau)$ the spin contribution would lead to 
$K_{\textrm{sympl}} (\tau) \ne K_{\CSE}(\tau)$.  This observation supports the 
hypothesis that the $\tau^m$-term derives only from diagrams with $m-1$ 
self-intersections.

\subsection{Spin rotations from subgroups of $\SU(2)$}\label{ss:subgroups}
If instead of the whole of $\SU(2)$ spin transformations are chosen from a 
subgroup $\Gamma$ it is still possible to find the connection between 
CSE and COE statistics.  Rather than averaging over $\SU(2)$ the identities
(\ref{identity 1}) -- (\ref{identity 3}) must be understood in terms of an 
average over $\Gamma$, ie
\begin{displaymath}
\int_{\SU(2)} f(u)\ \ud u \quad \textrm{is replaced by} \quad
\frac{1}{|\Gamma|} \sum_{g\in \Gamma} 
f(g)
\end{displaymath}
when $\Gamma$ is finite.
Let $\Gamma \subset \SU(2)$, viewed as a representation, be irreducible. 
The identities 
(\ref{identity 2}) and (\ref{identity 3}) can then be derived from Schur's 
lemma.  If and only if $\Gamma$ is also a quaternionic representation 
(any representation equivalent to its complex conjugate but inequivalent 
to any real representation)
the identity (\ref{identity 1}) may also be derived, 
see \cite{book:hamermesh}.  
As long as (\ref{identity 1}) -- (\ref{identity 3})
hold the argument is unaffected 
by the use of a subgroup of spin transformations.  CSE statistics hence depend
on the subgroup of spin transformations providing an irreducible quaternionic 
representation.

An example of a finite group of spin transformations 
are Hamilton's quaternions
\begin{equation}
\Gamma=\{ \pm \UI, \pm \ui \sigma_{x}, \pm \ui \sigma_{y},
\pm \ui \sigma_{z} \} \ ,
\end{equation}
where $\sigma_j$ is a Pauli matrix.
In \cite{paper:keppelermarklofmezzadri} spin transformations from this 
subgroup are applied to the cat map 
and CSE statistics  observed.  
As $\Gamma$ is both irreducible and quaternionic CSE statistics 
can indeed be expected with spin transformations taken even from
such a small subgroup of $\SU(2)$. 

The conditions (\ref{identity 1}) -- (\ref{identity 3}) depend only on the 
representation being irreducible and quaternionic consequently 
the argument also generalises 
to higher dimensional representations of $\SU(2)$, ie higher spins.  Let 
$\Gamma \subseteq \SU(2)$ and $R^s(\Gamma)$ an irreducible representation 
of dimension $2s+1$, ie a spin $s$ representation, then
\begin{equation}
\frac{1}{|\Gamma|}\sum_{g\in \Gamma} \tr \left( X \, \big( R^{s}(g) \big)^2 
\right) = 
\frac{c}{2s+1} \tr(X) \ .
\end{equation}
Here $c=1$ for real representations and $c=-1$ for quaternionic 
representations,
however only quaternionic representations can have even dimension
\cite{book:hamermesh}.  Therefore
$c=-1$, compare (\ref{identity 1}), implies $s$ half-integer.  
(\ref{identity 2}) and (\ref{identity 3}) generalise similarly.  
The RMT relation between 
the symplectic and orthogonal form factors (\ref{form factor relation})
can hence be derived for half-integer spin provided 
spin-transitions generate a 
quaternionic irreducible representation of $\Gamma$.


\subsection*{Acknowledgments}
We would like to thank Jonathan Robbins and Holger Schanz for their comments.
This work has been fully supported by the European Commission under the 
Research Training Network (Mathematical Aspects of Quantum Chaos) 
no. HPRN-CT-2000-00103 of the IHP Programme.

\bibliography{../../reffs/papers.bib,../../reffs/books.bib}

\end{document}